\documentclass[fleqn,usenatbib]{mnras}

\usepackage{newtxtext,newtxmath}

\usepackage[T1]{fontenc}

\DeclareRobustCommand{\VAN}[3]{#2}
\let\VANthebibliography\thebibliography
\def\thebibliography{\DeclareRobustCommand{\VAN}[3]{##3}\VANthebibliography}
\usepackage{xcolor}

\newcommand{\tamas}[1]{{#1}}

\usepackage{graphicx}	
\usepackage{amsmath}	
\usepackage{supertabular}
\usepackage{tablefootnote}

\title[CoRoT-TESS triples]{CoRoT-TESS eclipsing binaries with light-travel-time effect}

\author[T. Hajdu et al.]{
T. Hajdu$^{1,2,3,4}$,  
B. Mat\'{e}csa  $^3$, J. M. Sallai$^3$
and A. B\'{o}di$^{1,2,4}$\\
$^{1}$Konkoly Observatory, Research Centre for Astronomy and Earth Sciences (ELKH), H-1121 Budapest, Konkoly Thege Mikl\'os \'ut 15-17, Hungary\\
$^{2}$CSFK, MTA Centre of Excellence, H-1121 Budapest, Konkoly Thege Mikl\'os \'ut 15-17, Hungary\\
$^{3}$E\"{o}tv\"{o}s Lor\'{a}nd University, Department of Astronomy, H-1118 P\'{a}zm\'{a}ny P\'{e}ter stny. 1/A, Budapest, Hungary  \\
$^{4}$MTA CSFK Lend\"ulet Near-Field Cosmology Research Group, H-1121, Budapest, Konkoly Thege Mikl\'os \'ut 15-17, Hungary}

\date{Accepted XXX. Received YYY; in original form ZZZ}

\pubyear{2022}

\begin{document}
\label{firstpage}
\pagerange{\pageref{firstpage}--\pageref{lastpage}}
\maketitle

\begin{abstract}
Identifying long-period eclipsing binaries with space-based photometry is still a challenge even in the century of space telescopes due to the relatively short observation sequences and short lifetime of these missions. The Transiting Exoplanet Survey Satellite (TESS) space telescope is an appropriate tool to supplement previous space-based observations. 
In this paper we report the first results of the eclipse timing variation (ETV) analyses of eclipsing binaries (EBs) measured by CoRoT and TESS space telescopes. 
Among the 1428 EB candidates we found 4 new potential triple candidates, for which ETV was analysed and fitted by the well-known light-travel-time effect (LTTE). 
One of them shows significant phase shift in its folded light curve which required extra care.
In this paper we also present some other systems showing significant ETV signals that could be explained by mass transfer or apsidal motion.
\end{abstract}


\section{Introduction}

A large fraction of the detectable period changes observed in binary stars is caused by a third body, forming a gravitationally bound triple system. The evolution of these hierarchical systems can lead to the formation of exotic systems such as blue stragglers 
\citep{Naoz_Fabrycky2014} or binary neutron stars \citep{Shappee_Thompson2013}. Thus, there is no question that the study of triple systems is important to better understand the evolution of stars in stellar systems.


In the last decade, the discovery of triple stellar systems has become relatively easy thanks to the space-based photometry of \textit{Kepler} \citep{2014Conroy, 2015Borkovits, 2016Borkovits} and TESS \citep{2020Mitnyan, 2021Mitnyan, 2022Rappaport} space telescopes. While some of these systems were discovered due to the extra eclipses of the third body, the majority were detected by eclipse timing variation (ETV). Typically, the period of the outer orbit ranges from about a few months to several years. Nowadays, thanks to continuous and precise photometry, even really compact systems with periods less than a year have become easy to identify. In addition, the combination of these observations taken at different times is also promising for discovering longer outer period systems ($P_2\sim$ few years).

The first space-based telescope to collect long continuous, well-sampled, high-precision photometry for a large number of stars was CoRoT, whose data formed the basis of many pioneering works \citep{2009A&A...493..193L, 2009A&A...506..287L, 2010A&A...519A..88A, 2015A&A...576L..12C, 2015A&A...577A..11M}. 
CoRoT, similar to the \textit{Kepler} space telescope, observed thousands of eclipsing binaries, of which a comprehensive study has not yet been conducted.
Only a single study was based on CoRoT measurements that identified a few close hierarchical multiple systems \citep{2017Hajdu}. Unfortunately, due to the relatively short observation sequences, the generated ETVs alone are typically not suitable for finding multiples. However, based on the comprehensive analyses of the Kepler triples \citep{2016Borkovits}, there must be many more multiple system candidates but with longer outer periods. To find these systems the TESS space telescope can help thanks to its short but recurrent observations.

In this paper we present the first ETV analyses of binaries observed by both CoRoT and TESS space telescopes \citep{corot2009,ricker2015}. This is also the first time when we are able to identify such long (several years) period triple systems with the analyses of space telescopes measurements that cannot be discovery using only one observation sequence.

This paper is structured as follows. We formulate the basic mathematical background of the  third-body affected ETV analysis in Section \ref{LTTE}.
In Section \ref{data_preparation}, we outline the steps of our study, from selecting the systems and  data preparation to determining the orbital parameters of the candidates.
The results of the ETV analysis are discussed in Section \ref{results}, where we also present some other systems with significant ETVs.
Finally, a short summary is given in Section \ref{summary_and_conclusion}.

\section{Light-Travel-Time effect}\label{LTTE}

The analysis of hierarchical triple stellar systems plays a significant role in the  understanding of the evolution of short periodic binary systems \citep{Toonen2016}. These systems are basically studied by eclipse timing variation (ETV) which tells fundamental information about the outer orbit.
The light-travel-time effect (LTTE) that causes the main variation in the ETV was described by \citet{Irwin1952} in the following form:

\begin{equation}
    \Delta_\mathrm{LTTE}=-\frac{a_\mathrm{AB}\sin i_2}{c}\frac{\left(1-e_2^2\right)\sin\left(v_2+\omega_2\right)}{1+e_2\cos v_2},
\end{equation}
where $a_\mathrm{AB}$ denotes the semi-major axis of the EB's center of mass around the center of mass of the triple system, while $i_2$, $e_2$, $\omega_2$ and $v_2$ stand for the inclination, eccentricity, and argument of periastron of the relative outer orbit and the true anomaly of the third component, respectively. 
By these orbital parameters, the minimum mass of the third component can also be estimated through the  mass function, iteratively, if we know the mass of the eclipsing binary ($m_{EB}$):

\begin{equation}
\label{eq:Mass_function_eq}
f(m_\mathrm{C})=\frac{m_C^3\sin^3i_2}{m^2_\mathrm{ABC}}=\frac{4\pi^2a^3_\mathrm{AB}\sin^3i_2}{GP_2^2},
\end{equation}
where $m_C$ is the mass of the tertiary component and $m_{ABC}$ is the total mass of the triple system ($m_C+m_{EB}$).

In addition to the LTTE, dynamic effects can also be significant which was described in detail in a series of papers by \citet{Borkovits2003,Borkovits2011,2015Borkovits}. However, this only makes a significant contribution in really tight systems. It is typically taken into account when the $P_2/P_{EB}$ ratio is 100 or less. 

The minimum times can also be affected by other effects like stellar spots \citep[see e.g.,][]{Balaji_Kepler_spots}, apsidal motion \citep{1983Gimenez} and mass transfer which are not always distinguishable from the light-time effect, especially in case of inadequate sampling.

\section{Data preparation and analysis}\label{data_preparation}

\begin{figure}
\includegraphics[width=\linewidth, ]{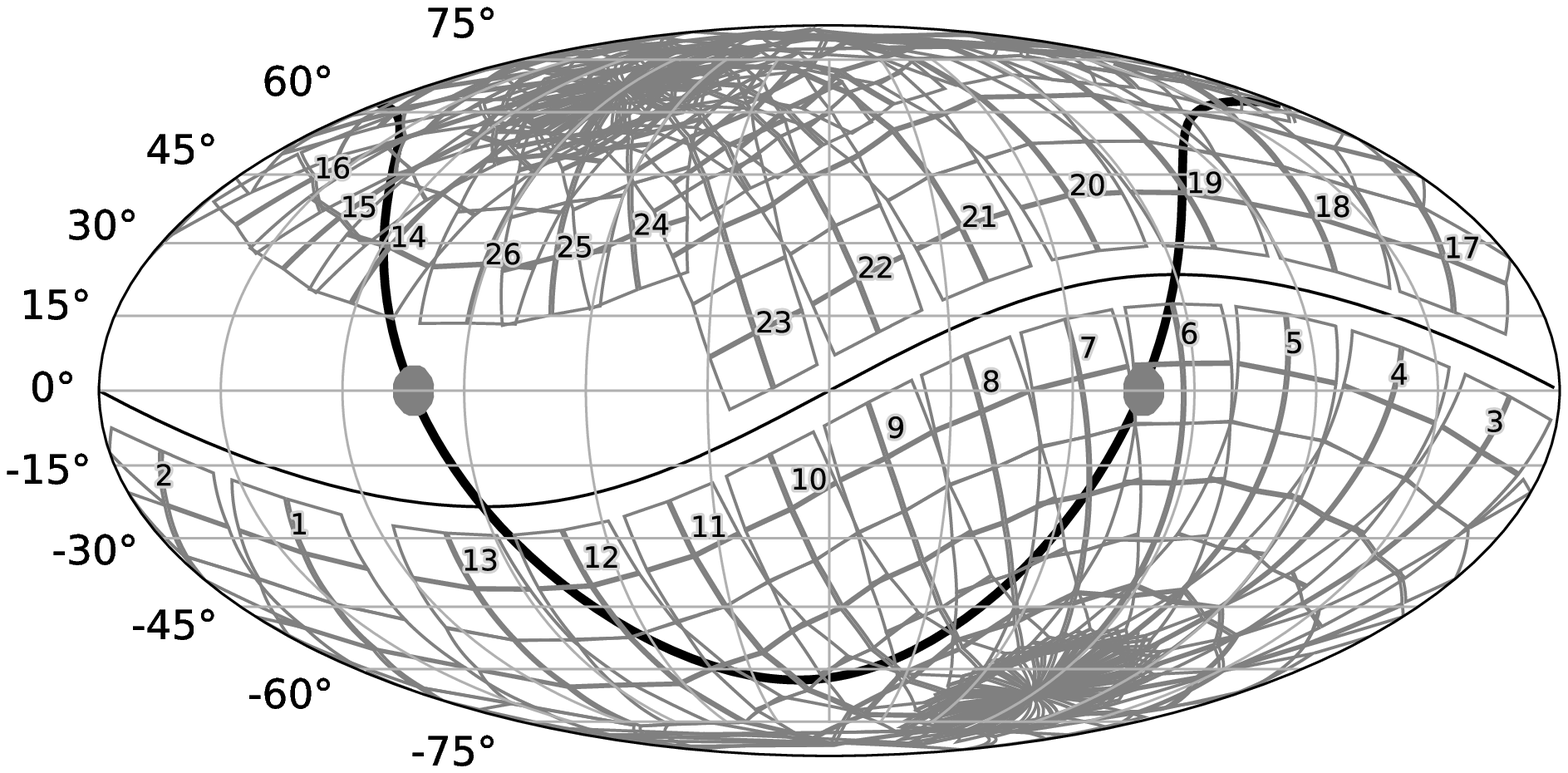}
\includegraphics[width=\linewidth, ]{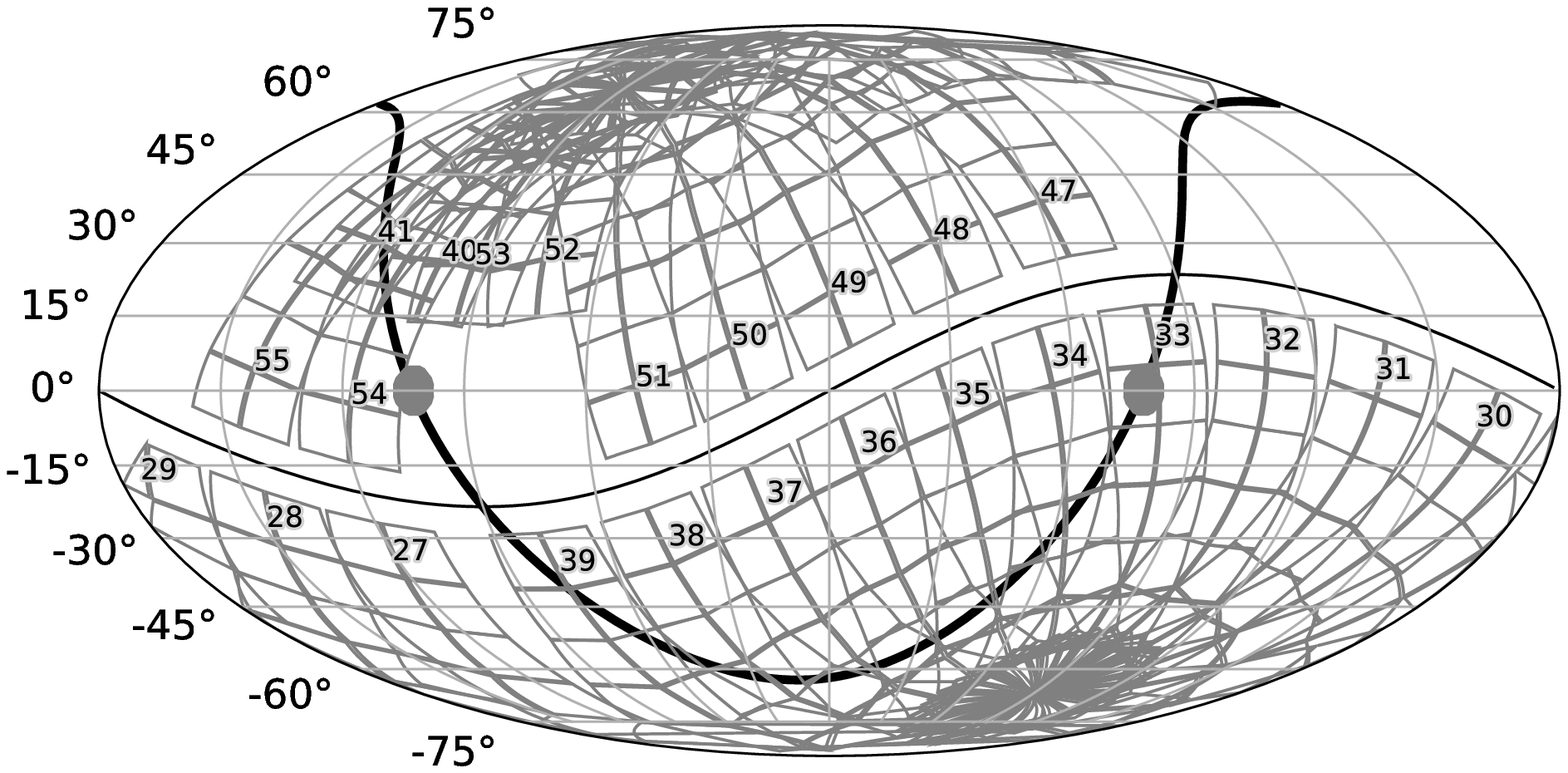}
\caption{The CoRoT (gray patches) and TESS fields of view in celestial coordinates. The numbers are the TESS sectors. The top panel represents the first 2 years of TESS measurements and bottom panel shows the following 2 years. The thick black line is the galactic plane. }
\label{CoRoT-TESS_sky}
\end{figure}

For our investigation we used the CoRoT EB candidate catalog published by \citet{2017Klagyivik} and the unofficial online CoRoT catalogue\footnote{\url{http://www.astro.tau.ac.il/~jdevor/CoRoT_catalog/catalog.html}} assembled by J.~Devor. From these catalogs we selected those targets which were also observed by TESS multiple times. Figure \ref{CoRoT-TESS_sky} shows the CoRoT (gray patches) and TESS fields of view in celestial coordinates. From this figure it can be seen that the direction of the galactic anticenter is the area which is covered by TESS sectors 6, 7 and 33, making the identification of triple candidates easier.


\begin{figure*}
    \centering
    \includegraphics[width=\columnwidth, ]{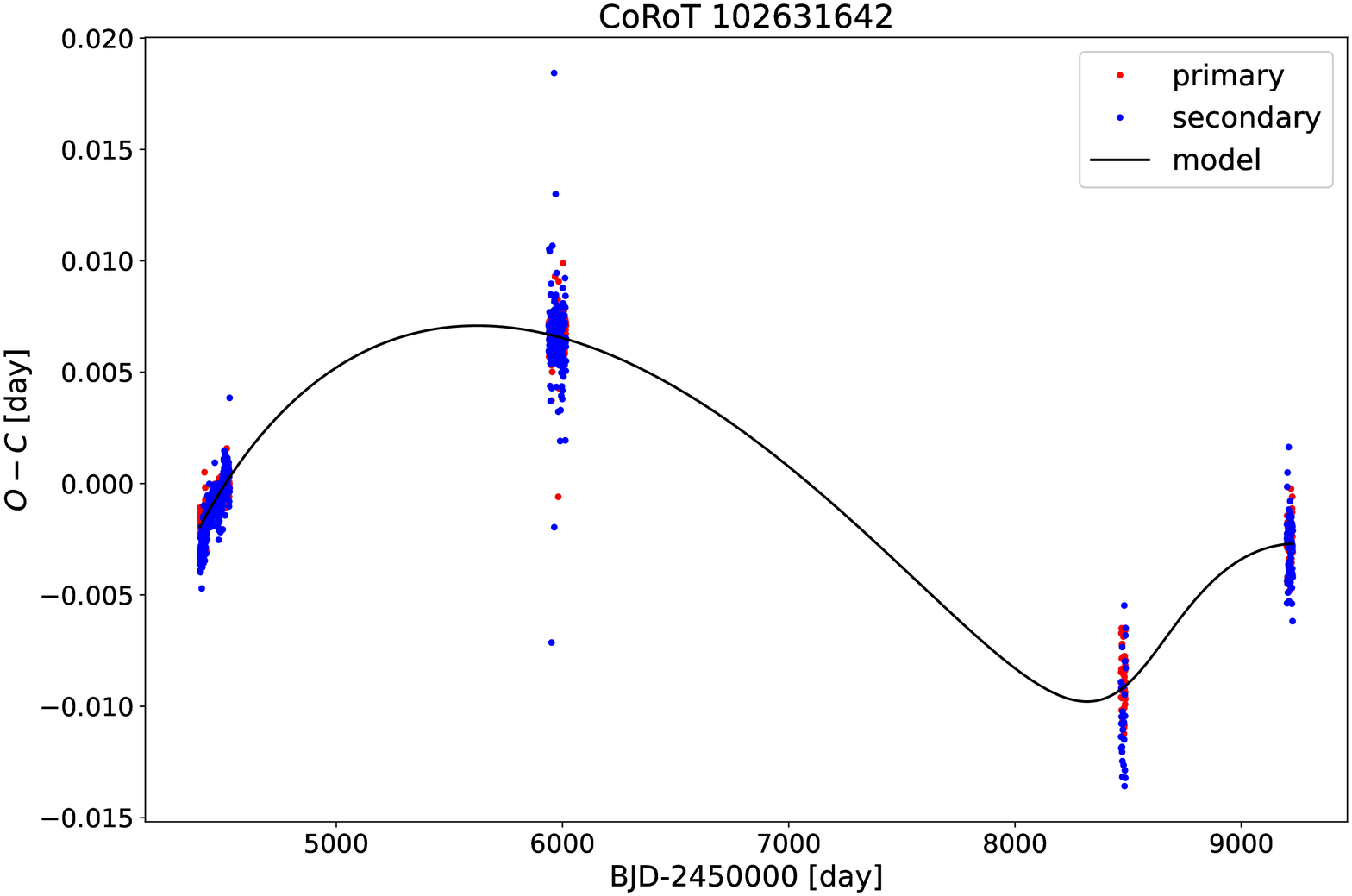}
    \includegraphics[width=\columnwidth, ]{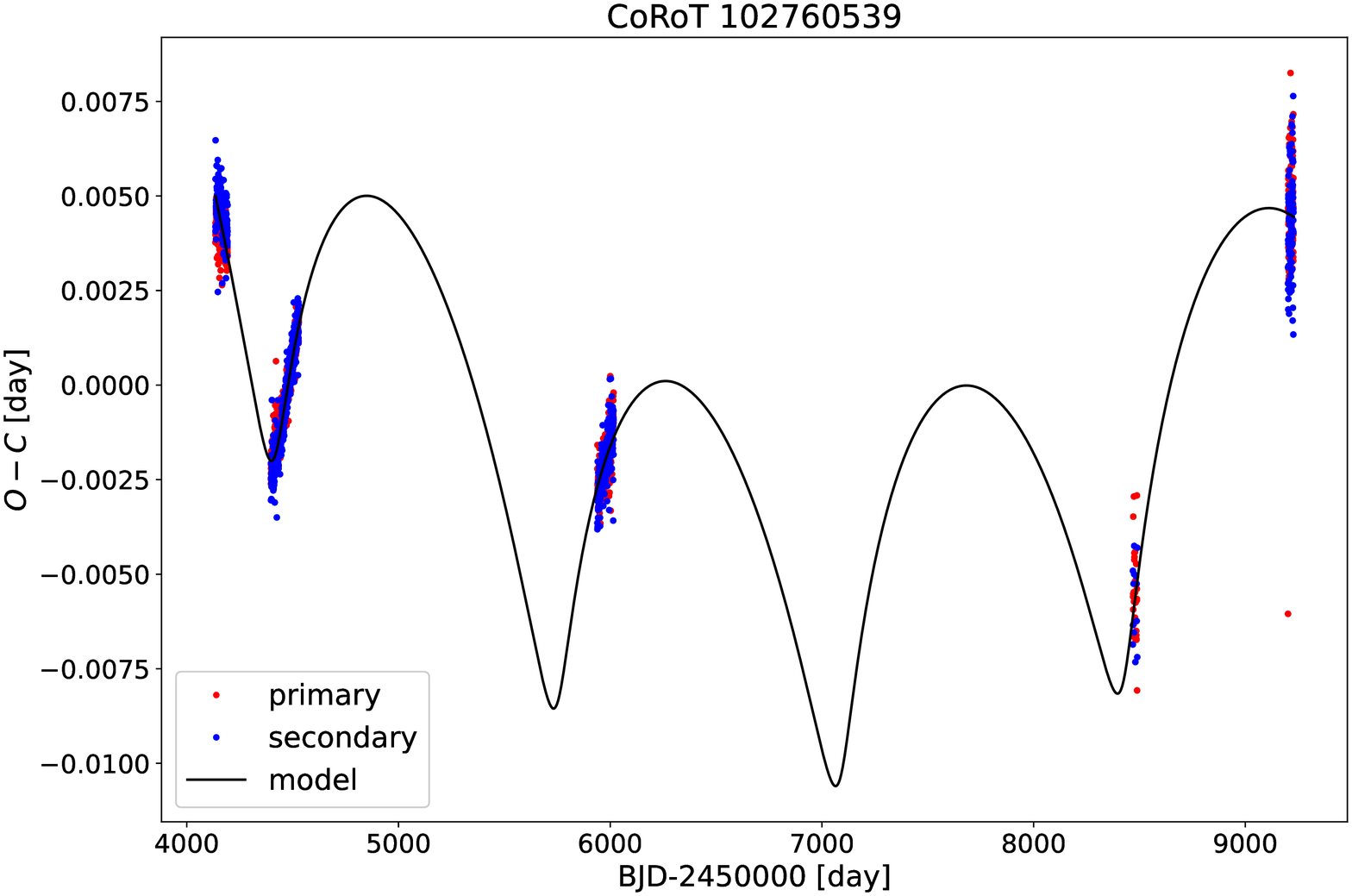}

    \includegraphics[width=\columnwidth, ]{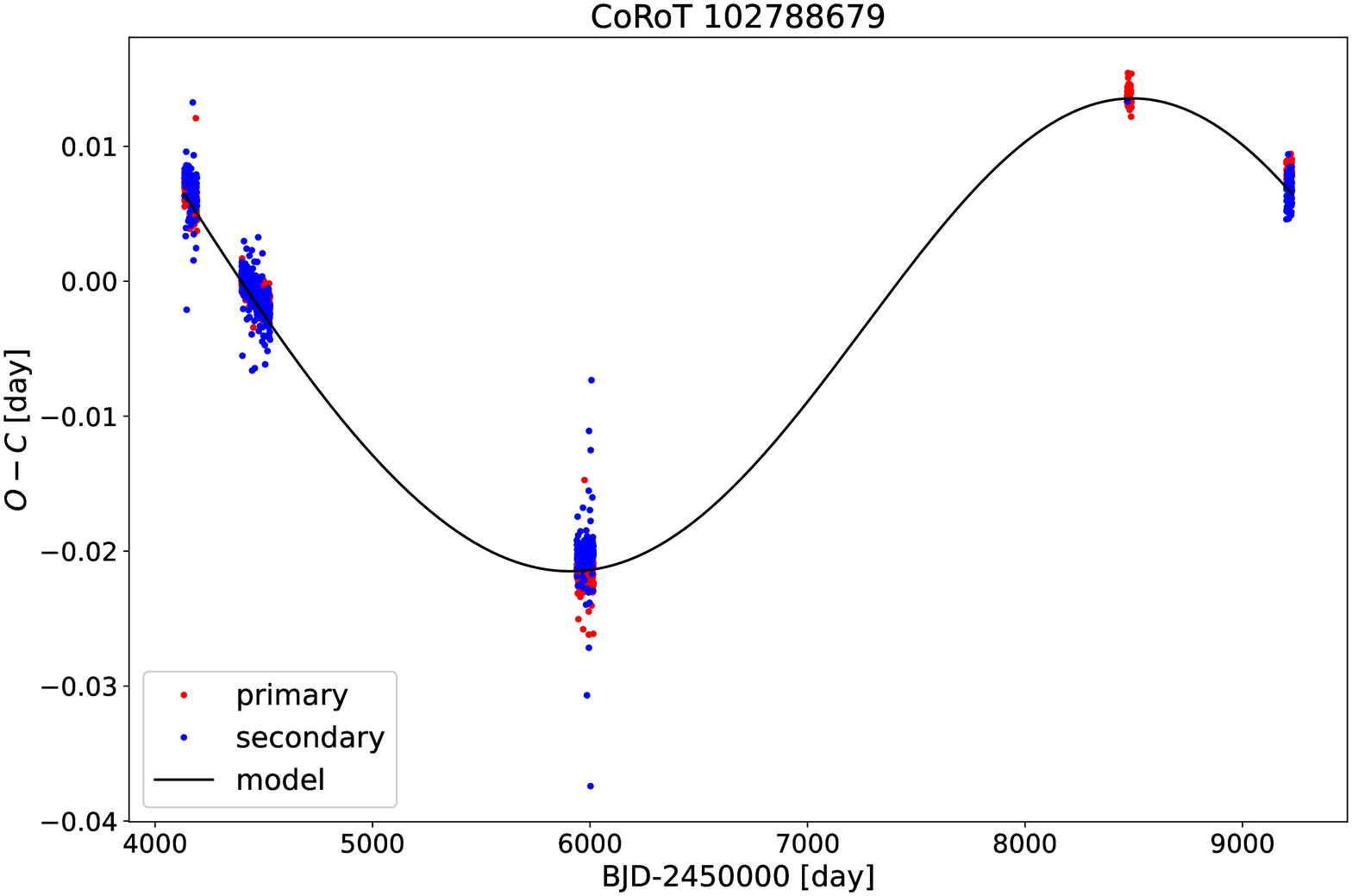}
    \includegraphics[width=\columnwidth, ]{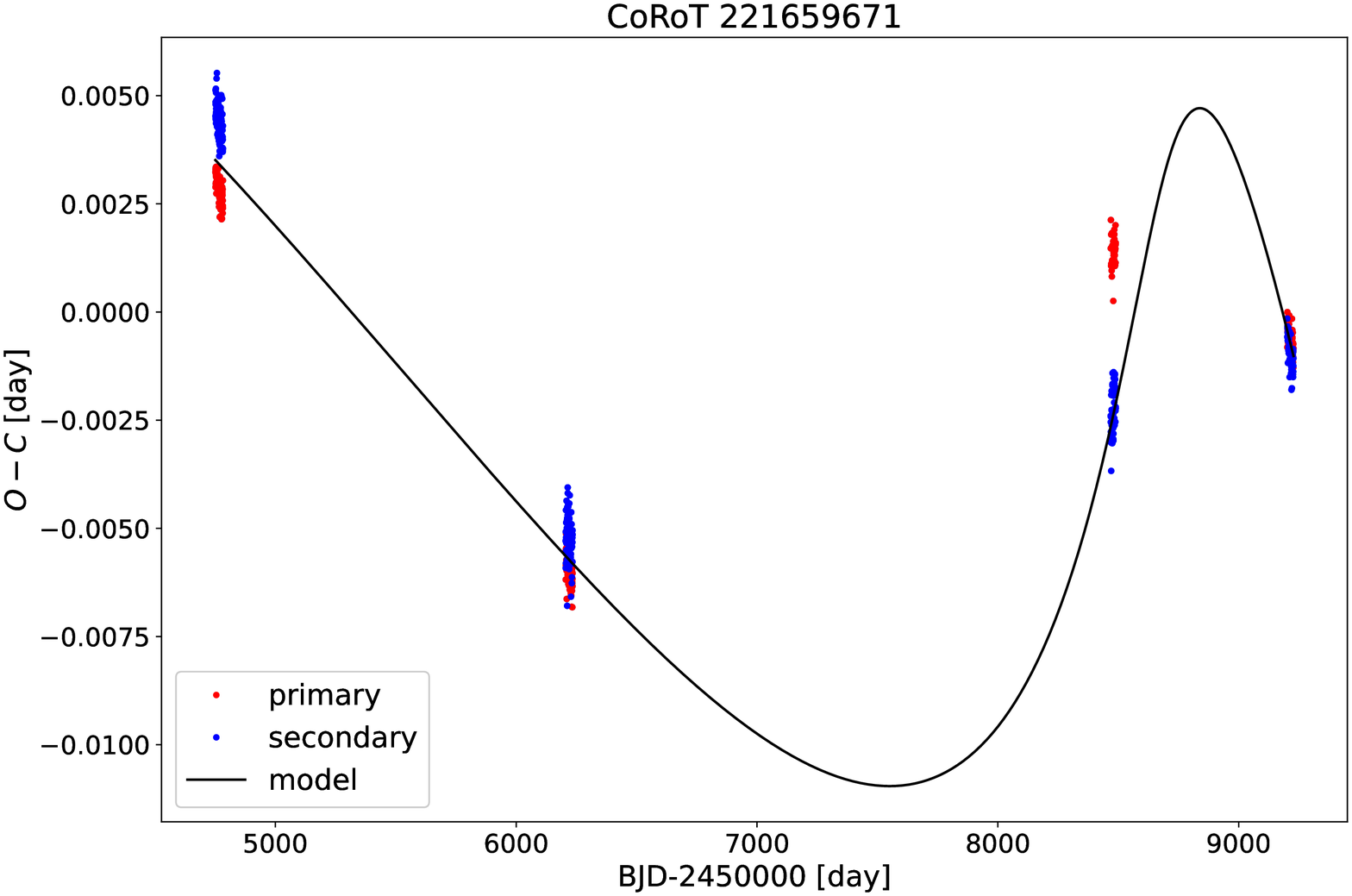}
    \caption{ETV of the systems and the best-fitting LTTE models (black curves). The red and blue points correspond to the primary and secondary minima, respectively.}
    \label{fig:all}
\end{figure*}

In the first step we determined the orbital period of the binaries with Phase Dispersion Minimization (PDM; \citealt{Stellingwerf78}) using the \texttt{cuvarbase}\footnote{\url{https://github.com/johnh2o2/cuvarbase}} python package. We used the so-called binned linear interpolation method with 100 bin cells. For this purpose we only used the CoRoT light curves.

In the next step we used the \texttt{TESScut} package \citep{Brasseur2019} to extract TESS light curves. For each target we downloaded the pixels within a $5\times5$ pixel area around the centroid. To build an aperture we selected only those pixels where the folded light curve had a signal-to-noise ratio (S/N) higher than 1. To always get a light curve, at least the pixel with the highest S/N value was selected. The value of the S/N was calculated from the parameters of the folded and binned light curve, namely the standard deviation of each bin and the amplitude of the binned light curve. To estimate the optimal number of bins, we used the same method as \citet{2021Bodi}.

Then, to remove the systematic trends we used the \texttt{flatten} method from the \texttt{wotan} python package \citep{2019Hippke}. We set the window length to twice the binary period. Furthermore, we cleared the light curves by removing the outlier points based on the standard deviation of each bin of the folded light curves. Finally, we rescaled the flux values of each observation series so their folded and binned light curves have the same amplitude. 
This step was necessary to find the correct orbital period of the systems and made it easy to identify systems  where the ETV caused dramatic apparent period changes (see e.g., the case of CoRoT 102788679 in section \ref{sec:102788679}). 

After this pre-processing, the period was recalculated with the full light curve. 

The ETV analysis was performed automatically using the method developed by \citet{2022Hajdu}. Then, because of the relatively low number of systems and the poor coverage of the outer orbits, the hierarchical triple candidates were selected manually.
We paid special attention to the systems that have been measured more than 3 times, as only under these conditions can the presence of a third body be determined with sufficient certainty. For the others, only the presence of a quadratic trend or apsidal motion could be determined.
The manual selection was also supported by the folded light curves, of which some showed significant phase shift.

For the selected candidates the ETV curves were fitted by LTTE, whose parameters were optimized with Levenberg-Marquardt algorithm. In order to get reliable errors we used the Markov chain Monte Carlo (MCMC) method as it is implemented in the \texttt{emcee} \citep{2013Foreman-Mackey} python package, similarly to \citet{2022Hajdu}. 

To calculate the minimum mass of the third body, we need to know the total mass of the inner binary. This can be estimated using the period-mass relation of the W~UMa-type binaries published by \citet{Dimitrov2015}.
Based on their paper the mass of the binary is:
\begin{equation}
    m_{AB}=\frac{0.0134}{P^2}\left(-1.154 +14.633\cdot P-10.319\cdot P^2\right)^3,
    \label{kettostomege}
\end{equation}
where $P$ stands for the orbital period of the binary. Using this formula and the equation of the mass function (Eq. \ref{eq:Mass_function_eq}) we were able to estimate the minimum mass of the third companions.

\begin{table*}
    \caption{The orbital parameters derived from Levenberg-Marquardt fit and their errors estimated via MCMC sampling.}
    \label{tab:parameters}
\begin{center}
    \begin{tabular}{ccccccccccc}
        \hline
          {CoRoT ID} & {$P_{EB}$} & $t_0$  & {$a_2\sin(i_2)$}& {$P_2$} & {$e_2$} & {$\tau_2$} & {$\omega_2$} & $f(m_c)$ & $M_{min}$\\
         & {[day]} & {[BJD-2450000 day]} & {[$R_\odot$]} & {[day]} & & {[BJD-2450000 day]}  &{[deg]} & {[$M_\odot$]} & {[$M_\odot$]} \\[0.2cm]
         \hline
         102631642 & $0.343116$ & $4397.3067$ & $ 304.81_{-41.26}^{+247.63} $ &$ 4567.11_{-811.71}^{+996.44} $ &$ 0.53_{-0.44}^{+0.20} $ &$ 8569.06_{-672.27}^{+1297.70} $ &$ 2.60_{-0.62}^{+1.41} $ & $ 0.02_{-0.01}^{+0.13}$ & $0.50_{-0.10}^{+0.65}$ \\[0.2cm]
         102760539 & $0.227558$ & $4134.9187$ & $194.73_{-11.10}^{+14.15} $ & $ 1336.37_{-4.37}^{+4.08} $ & $ 0.66_{-0.05}^{+0.05} $ & $ 4417.27_{-11.83}^{+9.23} $ & $ 2.00_{-0.10}^{+0.08} $ & $ 0.05_{-0.01}^{+0.02}$ & $0.54_{-0.07}^{+0.05}$ \\[0.2cm]
         102788679 & $0.243944$ & $4134.8952$ & $ 741.48_{-36.06}^{+580.10} $ & $ 5773.29_{-191.36}^{+1461.34} $ &
         $ 0.17_{-0.13}^{+0.03} $ & $ 7316.67_{-172.31}^{+713.37} $ & $ 3.44_{-0.40}^{+0.41} $ & $ 0.16_{-0.03}^{+0.39}$& $0.94_{-0.09}^{+0.70}$\\[0.2cm]
         221659671 & $0.292952$ & $4750.8828$ & $ 792.60_{-113.85}^{+292.45} $ &$ 5993.28_{-320.91}^{+536.90} $ &$ 0.55_{-0.14}^{+-0.01} $ &$ 8735.43_{-68.75}^{+295.72} $ &$ 3.83_{-0.16}^{+0.41} $ & $ 0.18_{-0.07}^{+0.32}$ & $1.16_{-0.22}^{+0.73}$\\[0.2cm]
         \hline
    \end{tabular}
\end{center}
\end{table*}

\section{Results}\label{results}
We have produced $O-C$ diagrams for 1428 binary systems, out of which we identified 4 potential hierarchical triple stellar system candidates. The orbital parameters of the systems derived from the LTTE fitting process are listed in Table \ref{tab:parameters}. The $O-C$ diagrams of the candidates and the best-fitting models are plotted in Figure \ref{fig:all}.

Because of the relatively low number of candidates we give some specific details about each of them below.

\subsubsection*{CoRoT 102631642}
CoRoT 102631642 is our only non-W~UMa-type binary system whose mass therefore cannot be determined based on its orbital period. Therefore, we used the usual approximation where the mass of the binary is assumed to be two solar masses. This system is more likely a $\beta$ Lyr system with the morphology parameter \citep{2008Prsa} $c=0.67$, which was calculated with the package published by \citet{2021Bodi}.

The $O-C$ of the system clearly shows a nearly sinusoidal LTTE (see top left panel of Fig. \ref{fig:all}). Nonetheless, the eccentricity of the outer orbit is large, and the fitted orbital parameters have the highest relative errors compared to the other systems in our list.

\subsubsection*{CoRoT 102760539}
CoRoT 102760539 is the only system that has a relatively short outer period ($\sim 1300^d$) and which could be identified as a triple without the TESS measurements. However, for the proper estimation of the orbital parameters, all measurements -- uniquely covering approximately 4 full outer orbits -- are necessary. Based on the period-mass correlation of W~UMa-type binaries \citep{Dimitrov2015}, the EB's mass was found to be $M_{EB}=1.14M_\odot$.

\subsubsection*{CoRoT 102788679}
\label{sec:102788679}
\begin{figure}
    \centering
    \includegraphics[width=\columnwidth]{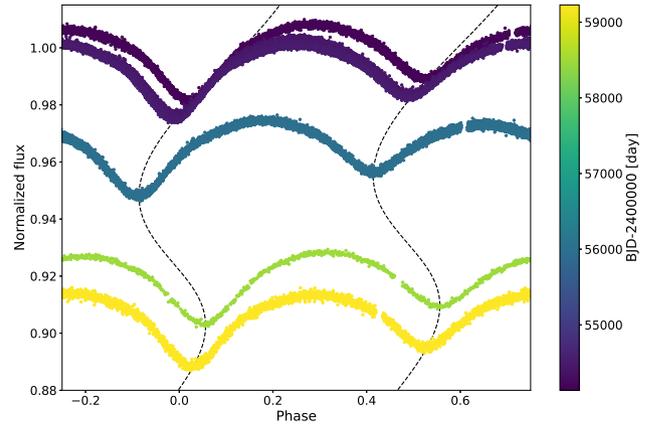}
    \caption{River plot of the folded light curve of CoRoT 102788679. For better visibility we shifted each observation sequence and the LTTE model (dashed lines) vertically as a function of the observation time, which is depicted by the color bar. 
    } 
    \label{fig:102788679folded}
\end{figure}

In most of the triple stellar systems with circular inner orbits the relative ETV ($dP/P_{EB}$) caused by the LTTE is small enough to be unnoticeable in the folded light curve. Only a few exceptions are known, like OGLE-BLG-ECL-253744 \citep{2022Hajdu} and KIC 9596187 \citep{2016Borkovits}. CoRoT 102788679 is a new member of this group, whose phase-folded light curve is shown in Fig. \ref{fig:102788679folded}, where we shifted each observation sequence and its LTTE model vertically according to the observation times to create a `river plot', similar to a waterfall diagram  \citep{Agol2018book}. \tamas{The folded light curve of the system also shows detectable O'Connell effect \citep{1951OConnell} which is most likely caused by starspots.}

Due to the relatively large $O-C$ amplitude, the PDM method, which was used to determined the orbital period, needed some manual intervention. Otherwise, it would have found a period by which in the folded light curve the two types of minima were interchanged. 

The $O-C$ shows the most sinusoidal-shaped LTTE among our targets, which is strengthened by the fitted small eccentricity value.

\tamas{In this type of active stars the angular momentum can also be disturbed by a magnetic torque which causes a nearly sinusoidal ETV. This effect is the so called Applegate mechanism \citep{1992Applegate} which may play a role in formation and evolution of close binaries. There are many papers which describe how the stellar parameters affect the period and the amplitude of the mechanism \citep{1998Lanza,2018Volschow} which may help to strengthen or disprove the presence of this effect.
}

\tamas{
We used the following formula to estimate the strength of the magnetic field \citep{2018Mitnyan}:}
\begin{equation}
    B^2 \sim 10\frac{GM^2}{R^4}\left(\frac{a}{R}\right)^2\frac{\Delta P}{P_{mod}},
\end{equation}
\tamas{
where $B$ is the strength of the magnetic field, $M$ and $R$ are the mass and the radius of a given component, respectively, $a$ is the semi-major axis, $\Delta P$ is the amplitude of the $O-C$ diagram and $P_{mod}$ is the period of the modulation. }

From the period-mass correlation, this binary system's mass was calculated to be $M_{EB}=1.3 M_\odot$. \tamas{Based on the Kepler's third law the semimajor axis is $a\sim 1.5R_\odot$. Also, using the relative parameters from the unofficial catalog of \textit{CoRoT} binaries\footnote{\url{http://www.astro.tau.ac.il/~jdevor/CoRoT_catalog/catalog.html}} 
we estimated the possible strength of the subsurface magnetic fields of both components separately. The parameters used to calculate the magnetic fields and the results are listed in Table \ref{tab:applegate}.  Based on these, if the secondary component has a magnetic field specified in the table, then the Applegate mechanism cannot be completely excluded from the possible causes of the period change. Similar example can be found in the case of VW Cep \citep{2018Mitnyan}.
}

\begin{table}
    \centering
    \caption{Stellar parameters of CoRoT 102788679. The first six values were used to estimate the components' subsurface magnetic field strengths. The last two rows are the results.}
    \begin{tabular}{l | c | l}
        Parameter & Value & Notes \\\hline
        $P_{EB}$ $[days]$ & 0.243944  & \\
        $M_{EB}$ $[M_\odot]$ & 1.3 &      Based on Eq.\ref{eq:Mass_function_eq}. \\
        $a$ $[R_\odot]$& 1.5 & Based on Kepler's third law.\\
        $M_2/M_1$ & 0.22 & Based on the unofficial catalog.\\
        $R_1/a$ & 0.313 & Based on the unofficial catalog. \\
        $R_2/a$ & 0.3949 & Based on the unofficial catalog. \\
        $B_1$ $[kG]$ & 129.1 & \\
        $B_2$ $[kG]$ & 15.3 & \\
    \end{tabular}
    \label{tab:applegate}
\end{table}

\subsubsection*{CoRoT 221659671}
This is the only system from our candidates where the light curve is significantly influenced by stellar spots. These spots also significantly affect the shape of the $O-C$ diagram, which can easily mislead the `normal' LTTE fitting process, especially if there is a large difference between the number of the data points 
in the two kinds of minima 
(primary and secondary)
. Therefore, for this system we calculated the average $O-C$ diagram, which not only smoothed out the effect of the spots, but also significantly reduced the number of data points to be fitted, which made the fitting process faster.

The fitted LTTE variation yielded the longest outer period -- about 6000 days -- among our targets. The mass of the binary was estimated to be $M_{EB}=1.8 M_\odot$. 

\subsection{Period-period distribution}
\begin{figure}
    \centering
    \includegraphics[width=\columnwidth]{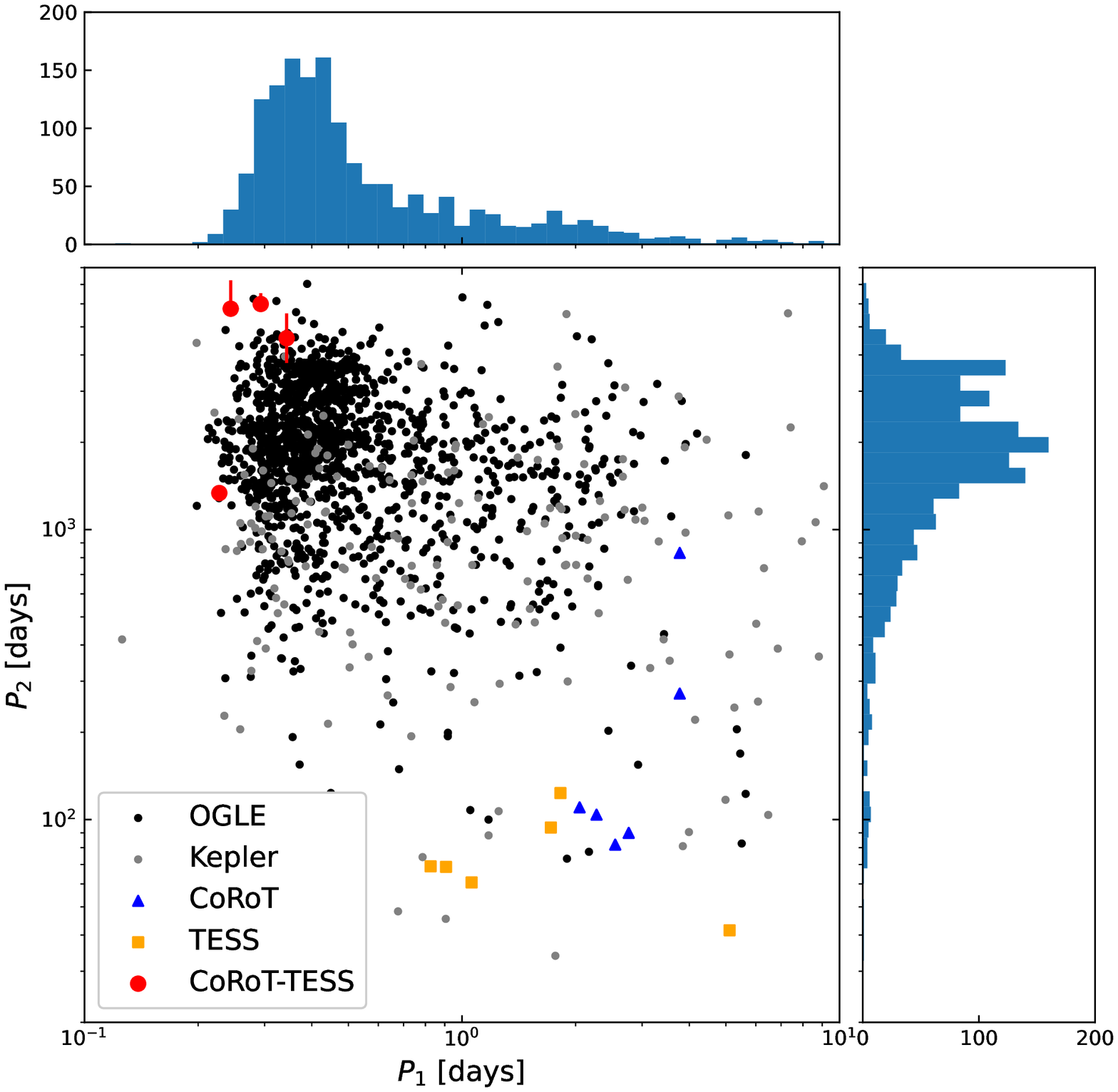}
    \caption{Period-period ($P_1-P_2$) diagram of triple systems. The black and gray points represent the OGLE \citep{2016Zasche, 2017Zasche, Hajdu2019, 2022Hajdu, 2022Hong} and the  \textit{Kepler}  \citep{2016Borkovits} triples, respectively, while the blue triangles and the orange squares stand for the \textit{CoRoT} \citep{2017Hajdu} and \textit{TESS} \citep{2022Borkovits, 2022Rappaport} hierarchical triple candidates, respectively. The red circles with the errorbars represent the four new \textit{CoRoT-TESS} triple stellar candidates. 
    }
    \label{fig:per-per}
\end{figure}

In Fig. \ref{fig:per-per} we show the period-period ($P_1-P_2$) distribution of triple systems, for which we used the data from several papers for comparison \citep{2016Borkovits,2022Borkovits,  2016Zasche, 2017Zasche,2017Hajdu,Hajdu2019, 2022Hajdu,2022Rappaport, 2022Hong}. The red dots with error bars are the four new CoRoT-TESS triple candidates. With the exception of CoRoT 102760539, all other candidates have longer outer periods than most triple systems and are on the edge of 
the clustering seen in OGLE systems. It can also be seen that the four new systems have longer outer period than any other CoRoT or TESS target.

\subsection{Systems with significant ETV} \label{sigETV}
During our manual search for triple stellar candidates we also found several other systems which show significant, but not strictly periodic ETVs. These systems and their basic parameters, the orbital period and epoch of the binary, and the type of the variation visible in their $O-C$ diagrams, are listed in Table \ref{tab:etv}. The full version of the table is available online. Based on the observed changes in the O-C diagram, 2 groups can be distinguished. 

\begin{table}
\centering
\caption{Further systems with significant ETV. The full list is available as online supplementary material.}
\label{tab:etv}
\begin{tabular}{c|c|c|c}
\hline
ID & $P_{EB}$ & $T_0$ & Type of variability  \\
& [day] & {[BJD-2450000 day]} & \\\hline
102285983	& $3.483354$	& $5108.164850$	&	parabolic ETV   \\   
102383485	& $7.830361$	& $5101.539991$	&	apsidal motion   \\   
102568782	& $0.589398$	& $4397.000981$	&	parabolic ETV   \\   
102573289	& $2.111683$	& $4395.621930$	&	apsidal motion   \\   
102574112	& $3.376890$	& $4396.478732$	&	apsidal motion   \\   

\multicolumn{4}{c}{...}\\\hline
    \end{tabular}
\end{table}

In the first group there are systems whose ETV shows parabolic trend which may be caused by mass transfer. However, we cannot rule out the possibility of the effect of a third body, but since we do not have enough observations, this cannot be decided.

\begin{figure}
    \centering
    \includegraphics[width=\columnwidth]{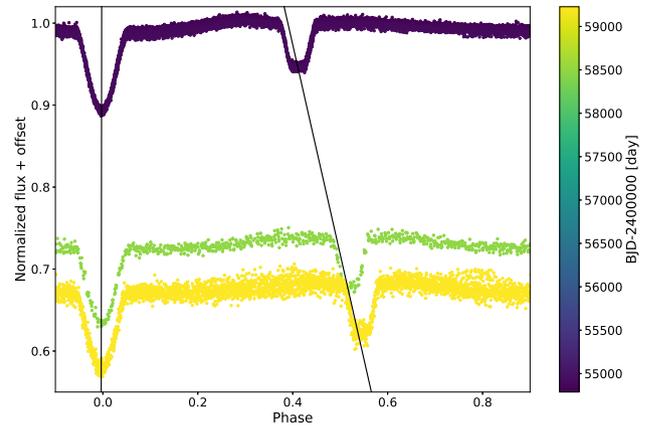}
    \caption{River plot of the folded light curve of CoRoT 110743209, where each observation sequence is shifted vertically according to its observation time. The solid black lines are only shown to highlight the variation of the eclipse midpoints.  }
    \label{fig:110743209}
\end{figure}

The second group contains systems whose $O-C$ diagrams show intercepting lines, which is the typical manifestation of apsidal motion. In some cases, the variation is also visible in the folded light curve, as well. A good example is CoRoT 110743209, whose folded light curve is plotted in Figure \ref{fig:110743209} and the primary minima are shifted to zero phase to highlight the variation in the secondary minima.

\section{Summary and conclusions}\label{summary_and_conclusion}

In this paper we report the results of a search for triple stellar candidates, which were observed by both the CoRoT and TESS space telescopes. To maximize the possibly of finding such systems, we pre-selected a list of targets that were observed in multiple TESS sectors. Out of about one and half thousand binary systems, we identified 4 new hierarchical triple candidates, whose eclipse timing variations (ETV) can be fitted with light-travel-time effect (LTTE). This number doubles the number of known CoRoT triples.

In case of each triple stellar system candidate, we determined and listed the periods of the inner binaries and the orbital parameters of the tertiary, along with its estimated mass. In three cases, only the TESS observations made it possible to discover the tertiary candidate. In one case, where the light curve also shows significant phase shift, the CoRoT observations were enough to reveal the variability, but they were not enough to determine the proper orbital parameters. This shows that the combination of different observation sequences made by different telescopes holds great potential for the study of multiple stellar systems.

Besides the triples, we also listed systems that show significant ETV, probably due to mass transfer between the components or apsidal motion, which could be the subject of a further research, possibly supplemented with ground-based follow-up observations.

\section*{Acknowledgements}
This paper includes data collected by the TESS mission. Funding for the TESS mission is provided by the NASA Science Mission Directorate. The CoRoT space mission was developed and operated by CNES, with contributions from Austria, Belgium, Brazil, ESA, Germany, and Spain. This project has been supported by the NKFIH-OTKA grant KH-130372 and the KKP-137523 `SeismoLab` \'Elvonal grant of the  Hungarian Research, Development and Innovation Office, by the Lend\"ulet Program of the Hungarian Academy of Sciences, project No. LP2018-7/2021, and by the MW-Gaia COST Action (CA18104). This research has made use of NASA’s Astrophysics Data System.

\section*{Data Availability}
The original CoRoT light curves are available at Vizier\footnote{\url{https://vizier.cds.unistra.fr/viz-bin/VizieR-3?-source=B/corot}}. The full table of systems with significant ETV is available online as an external material.
The derived data generated in this research will be shared on \url{https://konkoly.hu/KIK/data_en.html} web page. 

\bibliographystyle{mnras}
\bibliography{main} 






\bsp	
\label{lastpage}
\end{document}